\documentclass[pra,twocolumn,showpacs,preprintnumbers,floatfix,superscriptaddress]{revtex4}
\usepackage{amsmath}
\usepackage{amsfonts}
\usepackage{amssymb}
\usepackage{graphicx}
\usepackage{epsfig}
\usepackage{dcolumn}
\usepackage{bm}
\usepackage{relsize}
\usepackage[usenames,dvipsnames]{color}	 
\usepackage[colorlinks=true,breaklinks=true,bookmarksopen=true,bookmarksnumbered=true,citecolor=blue,linkcolor=red,hyperfootnotes=false]{hyperref}  
\usepackage{pxfonts}
\begin{document}
\title{Elastic electron scattering from Be, Mg, and Ca}
\author{Mehrdad Adibzadeh}
\email{madibzadeh@uwf.edu}
\affiliation{Department of Physics, University of West Florida, Pensacola, FL 32514, USA}
\author{Constantine E. Theodosiou}
\email[Corresponding author: ]{  constant.theodosiou@manhattan.edu}
\affiliation{Department of Physics and Astronomy, Manhattan College, Riverdale, NY 10471, USA.}
\author{Nicholas Harmon}
\email{nharmon@coastal.edu}
\affiliation{Department of Physics, Coastal Carolina University, Conway, SC 29528-6054, USA.}
\date{\today}

\begin{abstract}
We present a comprehensive set of theoretical results for differential, integrated, and momentum transfer cross sections for the elastic scattering of electrons by beryllium, magnesium and calcium, at energies below 1 keV. In addition, we provide Sherman function values for elastic electron scattering from calcium in the same energy range. This study extends the application of our method of calculations, already employed for barium and strontium, to all stable alkaline-earth-metal atoms. Our semi-empirical approach to treating target polarization has produced in our earlier work a satisfactory agreement with experimental values and precise theoretical results such as convergent close-coupling calculations for barium. The present data are expected to be of similar high accuracy, based on our previous success in similar calculations for barium and all inert gases. 
\end{abstract}

\maketitle

\section{Introduction}
In our previous paper \cite{Adibzadeh2004}, we extended our relativistic approach to treat elastic electron collisions with inert gases \cite{Adibzadeh2005} to two heavy alkaline-earth-metal atoms, barium and strontium. Our results compared well with the few available experimental and the convergent close-coupling (CCC) calculations data for barium, and hence we presented our set of values of cross sections for strontium. The present work intends to expand the data to all stable alkaline-earth-metal atoms.
Similar to strontium, there is a small amount of experimental data available for elastic electron scattering from beryllium, magnesium and calcium. In fact, the elastic electron scattering from neutral beryllium, to the best of our knowledge, has attracted no experimental attention at all. Therefore, all available data come from theoretical calculations. The introduction of beryllium as a surface material in the JET project created a need for atomic data in early and mid 1990s. This motivated some precise calculations in \textit{R}-matrix and CCC formulations \cite{Fon1992, Fursa1997a,Fursa1997b,Bartschat1996,Bartschat1997}. However, those calculations did not provide much information on elastic scattering. The theoretical data on elastic electron scattering form beryllium was for long (up until 2016) limited to some older calculations. Those included the relativistic calculations of Fink and Ingram \cite{Fink1972} for 100, 250, 500, 1000, and 1500 eV, the close-coupling calculations of Fabrikant \cite{Fabrikant1975} for energies below 8.16 eV, the nonrelativistic partial-wave calculations of Kaushik \textit{et al.} \cite{Kaushik1983} for a few energies in the 5-30 eV range, and the low-energy integrated cross section (ICS) calculations of Yuan and Zhang \cite{Yuan1989} for energies below 1 eV. The work of Fink and Ingram \cite{Fink1972} also gives the only data on the spin polarization of electrons scattered from the beryllium atom. Recent B-spline and CCCC calculations of Zatsarinny \textit{et al.}~\cite{Zatsarinny2016} for energies up to 100 eV and relativistic complex optical potential (ROP) calculations of McEachran \textit{et al.}~\cite{McEachran2018Be} for energies up to 5000 eV provided theoretical data with more precision and range to the field.

For magnesium, however, there have been a few experimental works on elastic scattering, most of which date back to the 1970s and early 1980s. Burrow and Comer \cite{Burrow1975} and Burrow \textit{et al.} \cite{Burrow1976} performed experiments to identify the low-energy resonant scattering using the electron transmission method. Kazakof and Khristoforov \cite{Kazakov1982} studied this $^{\text{2}} {\text{P}}$ resonance by measuring the differential cross section (DCS) at angle $90^ \circ$. Nevertheless, the only experimental DCS and ICS data for magnesium are those provided by Williams and Trajmar \cite{Williams1978} for 10, 20, and 40 eV and, more recently, Predojević \textit{et al.}~\cite{Predojevic2007} for energies up to 100 eV. On the theoretical side, there are three old close-coupling calculations for magnesium. Van Blerkom~\cite{VanBlerkom1970} reported a few low-energy ICS's and Fabrikant provided low-energy cross sections in Refs.~\cite{Fabrikant1975,Fabrikant1974} and ICS and DCS values for 10 and 20 eV in Ref.~\cite{Fabrikant1980}. Comparison between the ICS's of Van Blerkom~\cite{VanBlerkom1970} and Fabrikant~\cite{Fabrikant1974} yielded a total disagreement. Other theoretical investigations for magnesium are the relativistic treatment of Gregory and Fink~\cite{Gregory1974} for 100, 250, 500, 1000, and 1500 eV impact energies, nonrelativistic partial-wave calculations of Khare \textit{et al.}~\cite{Khare1983} for some intermediate energies, the low-energy DCS and ICS results of Yuan and Zhang~\cite{Yuan1989,Yuan1993}, coupled-channels approximation calculations of Mitroy and McCarthy~\cite{Mitroy1989}, nonrelativistic partial-wave calculations of Pandya and Baluja~\cite{Pandya2011}, B-spline calculations of Zatsarinny \textit{et al.}~\cite{Zatsarinny2009} and ROP calculations of McEachran \textit{et al.}~\cite{McEachran2018Mg} for energies up to 5000 eV. In the absence of experimental data, the work of Gregory and Fink~\cite{Gregory1974} is the only available study on the spin polarization of scattered electrons from magnesium, which provides Sherman functions at the aformentioned energies.

In the case of calcium, for long the only experimental work was that of Romanyuk \textit{et al.}~\cite{Romanyuk1980}, which provided low-energy ICS values and was refined some years later~\cite{Romanyuk1992}. The measurement of differential and integrated cross section values for energies up to 100 eV by Milisavljevi\'c \textit{et al.}~\cite{Milisavljevic2005} is the only other experimental work for elastic scattering of electrons off the calcium atom. The theoretical calculations for calcium include the close-coupling calculations of Fabrikant~\cite{Fabrikant1975} for energies below 5.44 eV, a relativistic study on some intermediate energies by Khare \textit{et al.}~\cite{Khare1985}, data by Gregory and Fink~\cite{Gregory1974} at the same abovementioned energies, the low-energy studies of Yuan and Zhang~\cite{Yuan1989,Yuan1993}, Cribakin \textit{et al.}~\cite{Cribakin1992}, Yuan~\cite{Yuan1995}, Yuan and Fritsche~\cite{Yuan1997} and Yuan and Lin~\cite{Yuan1998}, partial wave oriented calculations of Raj and Kumar~\cite{Raj2007}, B-spline and R-matrix studies of Zatsarinny \textit{et al.}~\cite{Zatsarinny2006} for energies up to 4 eV and later up to 100 eV~\cite{Zatsarinny2019},  and optical potential calculations of Wei \textit{et al.}~\cite{Wei2020}. On the subject of spin polarization of scattered electrons from calcium, there are three works in the literature, all from the theoretical sector. Gregory and Fink ~\cite{Gregory1974} and Khare \textit{et al.}~\cite{Khare1985} provide Sherman functions for some intermediate and high energies, whereas the Sherman function values of Yuan~\cite{Yuan1995} are limited to a few low impact energies.

As noted, there has been very little experimental attention toward these three atoms. The present work aims to furnish a consistent set of elastic DCS, ICS, and MTCS data in a wide energy range and for all these three atoms, which could be useful to future experimental works.

\section{Brief Review of the Theoretical and Computational Approach}
The method of calculations of the present work is the same as that described in our previous papers~\cite{Adibzadeh2004, Adibzadeh2005} and to avoid repetition we refer the reader to them. In summary, we followed the standard method of partial-wave expansion in potential scattering, where the phase shifts were obtained by solving the stationary Dirac equation. The choices for central static atomic, exchange, and polarization potential in this work are the same as those used in Ref.~\cite{Adibzadeh2004}. In addition to cross section values, with the relativistic treatment, we are also able to compute the \textit{Sherman function} (SF), which describes the measured spin-up and spin-down asymmetries in the number of scattered electrons~\cite{Sherman1956, Kessler1969}. The present work provides Sherman function values only for calcium. The static polarizabilities, used in the polarization potential, were taken from the theoretical values of Kolb \textit{et al.}~\cite{Kolb1982}. The cutoff radius $r_c$, which appears in the polarization potential, is again assumed as a function of the incident electron energy. The functional behavior of this energy-dependent cutoff radius was chosen to be similar to those already used for barium and strontium~\cite{Adibzadeh2004}. Therefore the cutoff radius was assumed to be (in atomic units) for beryllium

\begin{equation}
r_c (E) = \left\{ {\begin{array}{*{20}c}
   {\frac{1}{3}\ln (\frac{E}{\mathcal{R}}) + \left\langle r \right\rangle _{2s} }
   & {E > 30{\rm{ eV}}},  \\\\
   {{\rm{2}}{\rm{.75}}} & {E \leq 30{\rm{ eV}}},  \\
\end{array}} \right.
\end{equation}

for magnesium,

\begin{equation}
r_c (E) = \left\{ {\begin{array}{*{20}c}
   {\frac{1}{3}\ln (\frac{E}{\mathcal{R}}) + \left\langle r \right\rangle _{3s} }
   & {E \ge 20{\rm{ eV}}},  \\\\
   {{\rm{3}}{\rm{.1}}} & {E < 20{\rm{ eV}}},  \\
\end{array}} \right.
\end{equation}

and for calcium,

\begin{equation}
r_c (E) = \left\{ {\begin{array}{*{20}c}
   {\frac{1}{3}\ln (\frac{E}{\mathcal{R}}) + \left\langle r \right\rangle _{4s} }
   & {E \ge 10{\rm{ eV}}}.  \\\\
   {{\rm{3}}{\rm{.7}}} & {E < 10{\rm{ eV}}}.  \\
\end{array}} \right.
\end{equation}

Here $E$ is the energy of the incident electron in eV, $\mathcal{R}$ is the Rydberg 
constant ($\mathcal{R}$ = 13.605 691 72 eV), and $\left\langle r \right\rangle
_{2s} = \text{2.61 }a_0$,  $\left\langle r \right\rangle
_{3s} = \text{3.12 }a_0 $ and  $\left\langle r \right\rangle
_{4s} = \text{3.98 }a_0 $ are the expectation values of the beryllium's $2s$ shell, magnesium's $3s$ shell, and calcium's $4s$ shell radii, respectively. The cutoff radii for low energies are set to constant values to avoid the anomaly caused by the logarithmic term. For more on this constant value, the reader may consult with Ref.~\cite{Adibzadeh2004}. Throughout this work and for all considered energies, we used 150 partial-wave phase shifts, which ensured the last spin-up or spin-down phase shift to be less than $10^{-6}$.

\section{Results and Discussion}

\subsection{Beryllium}
\begin{figure*}
\includegraphics{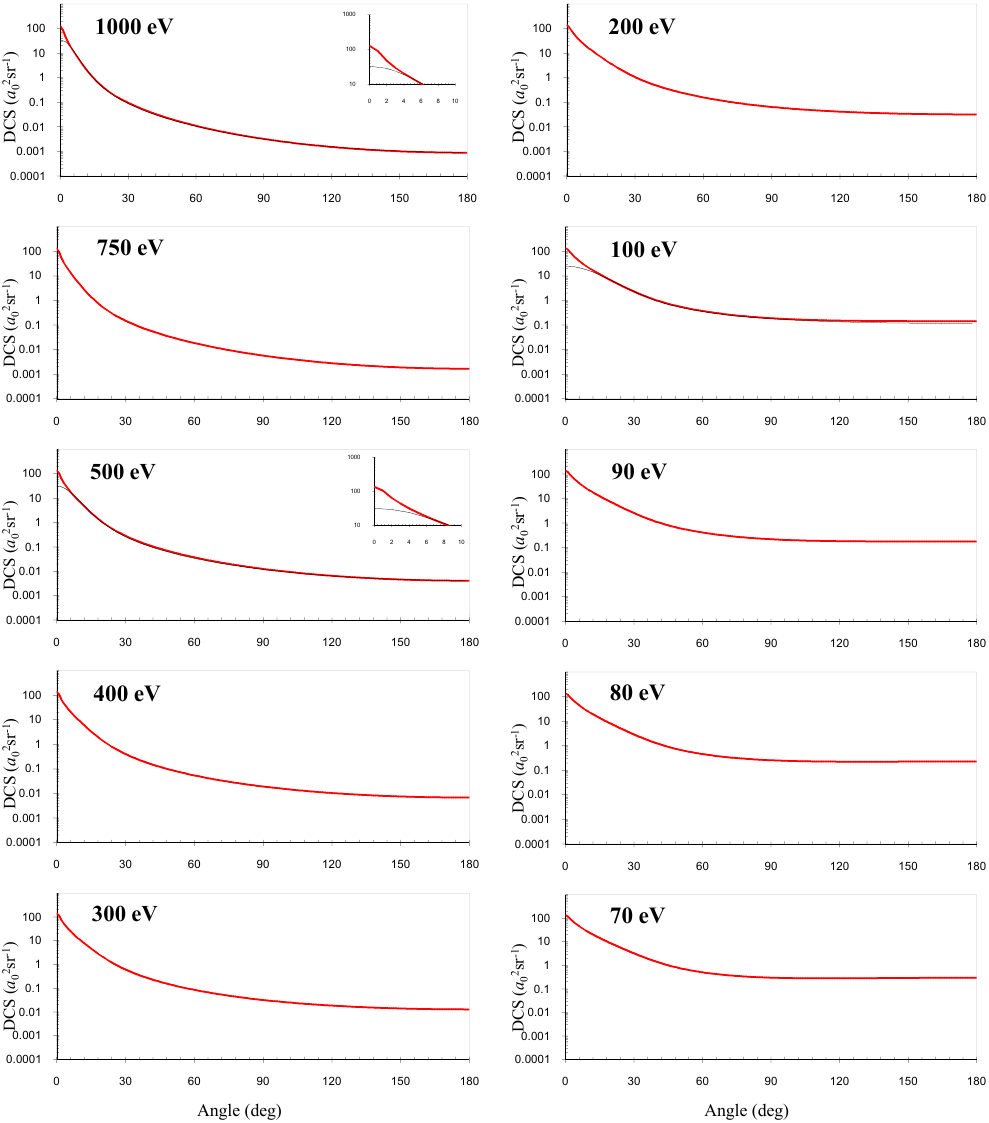}
\caption{\label{fig:be1} Differential cross sections for elastic electron scattering from beryllium at energies between 70 and 1000 eV: thick red solid line, present work; Other theoretical: thin black solid line, Fink and Ingram~\cite{Fink1972}.}
\end{figure*}

\begin{figure*}
\includegraphics{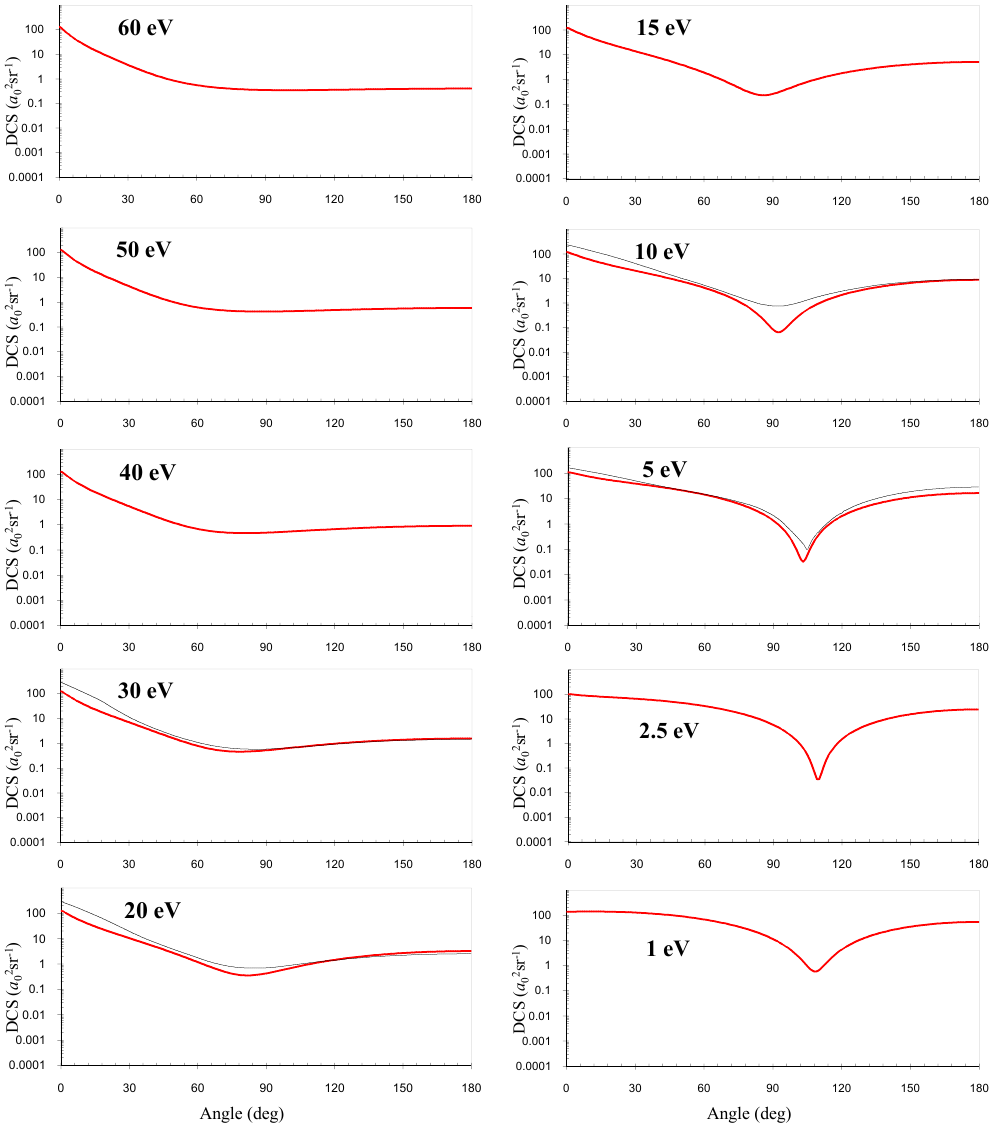}
\caption{\label{fig:be2} Differential cross sections for elastic electron scattering from beryllium at energies between 1 and 60 eV: thick red solid line, present work; Other theoretical: thin black solid line, Kaushik \textit{et al.}~\cite{Kaushik1983}.}
\end{figure*}

\begin{figure*}
\includegraphics{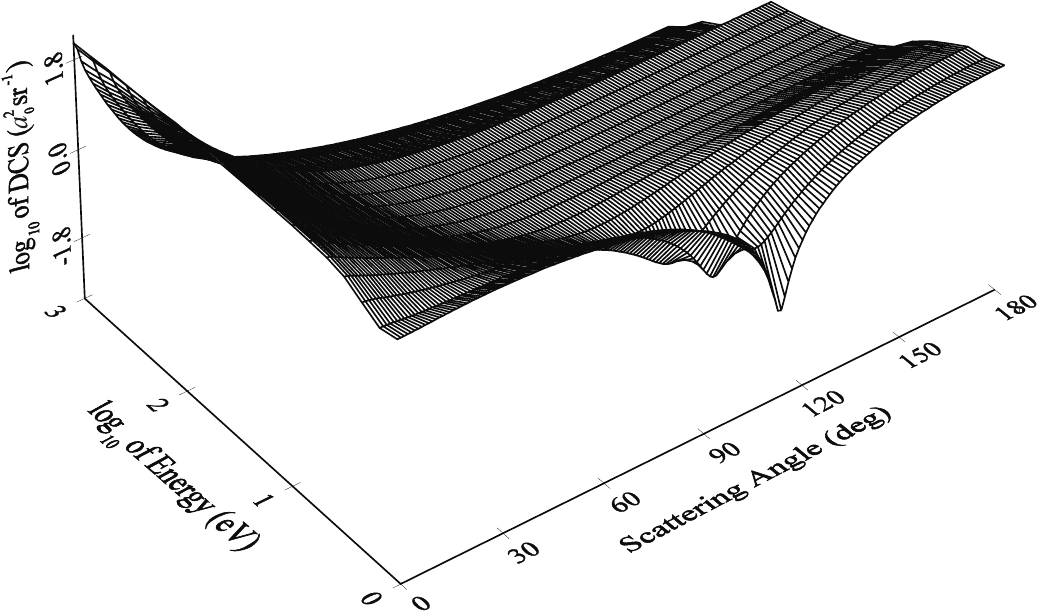}
\caption{\label{fig:be3} a three-dimensional view of differential cross section for elastic electron scattering from beryllium.}
\end{figure*}

\begin{figure*}
\includegraphics{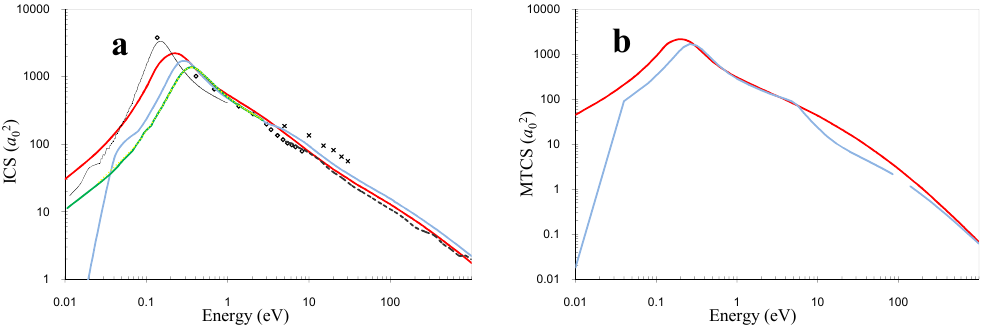}
\caption{\label{fig:be4} Integrated (\textbf{a}) and momentum transfer (\textbf{b}) cross sections for elastic electron scattering from beryllium: thick red solid line, present work; Other theoretical: thin black solid line, Yuan and Zhang~\cite{Yuan1989}; $\times$, Kaushik \textit{et al.}~\cite{Kaushik1983}; $\lozenge$, Fabrikant~\cite{Fabrikant1975}; dashed black line, Fursa and Bray\cite{Fursa1997a,Fursa1997b}; thick blue solid line, McEachran \textit{et al.}~\cite{McEachran2018Be}; thick green solid line, BSR-660 data of Zatsarinny \textit{et al.}~\cite{Zatsarinny2016}; thick dotted yellow line, CC-409 data of Zatsarinny \textit{et al.}~\cite{Zatsarinny2016}.}
\end{figure*}

Our DCS values for beryllium are shown in Figs.~\ref{fig:be1} and \ref{fig:be2}. As we already mentioned, there are no experimental data available to compare with. Our results, however, are in good agreement with the DCS values of Fink and Ingram~\cite{Fink1972} at 100, 500, and 1000 eV, except for the forward direction. There is a marked discrepancy between our DCS's and those of Fink and Ingram~\cite{Fink1972} at small scattering angles, as seen in Fig.~\ref{fig:be1}. This is understood, knowing that the calculations of Fink and Ingram~\cite{Fink1972} did not account for the effect of polarization of the atomic target and, less importantly, the effect of exchange in the scattering process. This difference is limited to the very forward direction angles at high energies, as can be seen in the magnified portions of 500 and 1000 eV graphs, but expands to larger angles at lower energies, e.g. 100 eV. This behavior is expected since the polarization of the atomic target contributes into a wider scattering angle range, in the forward direction, when interacting with slower incident electrons. The comparisons in Fig.~\ref{fig:be2} show a behavioral agreement between our DCS's and those of Kaushik \textit{et al.}~\cite{Kaushik1983}. At 30 and 40 eV, our DCS is visibly lower than those of Kaushik \textit{et al.}~\cite{Kaushik1983} and at 10 and 5 eV disagreements on the position of the DCS's minimum are more pronounced. To visualize the behavior of differential cross section as a function of impact energy and scattering angle, we present in Fig.~\ref{fig:be3} a three-dimensional (3D) graph of the logarithm ($\log _{10}$) of beryllium's DCS versus logarithm of the energy and the scattering angle. The simplicity of the atomic structure of the beryllium atom is even more obvious if one compares Fig.~\ref{fig:be3} with the 3D graphs presented in Refs.~\cite{Adibzadeh2004, Adibzadeh2005} for heavier atoms. Our elastic ICS and MTCS values for beryllium and comparisons with four other theoretical results are shown in Fig.~\ref{fig:be4}.

Our ICS values are in agreement with the results of CCC106 calculations by Fursa and Bray~\cite{Fursa1997a,Fursa1997b} over the energy range 5-1000 eV. The same agreement exists with B-spline and CCCC values of of Zatsarinny \textit{et al.}~\cite{Zatsarinny2016} and ROP values of McEachran \textit{et al.}~\cite{McEachran2018Be} down to around 0.5 eV. However, our ICS values attain a different maximum than those of other theoretical results at low energies. Our ICS display a maximum at higher energy than those of Yuan and Zhang~\cite{Yuan1989} and Fabrikant \cite{Fabrikant1975} by about 0.1 eV while being at lower energy than those by Refs.~\cite{McEachran2018Be} and \cite{Zatsarinny2016} by about the same 0.1 eV. From this maximum point down to 0.01 eV, the disagreement between the various theoretical results is rather profound. In the absence of experimental ICS data, no solid conclusion can be drawn in this region.  However, the agreement with other theoretical calculations is a very credible test for our ICS values at energies above 0.5 eV. Unfortunately, the momentum transfer cross section has not attracted the same theoretical attention. There is only one case of MTCS theoretical data available~\cite{McEachran2018Be} with which our values agree fragmentaly over the wide energy range of 0.5-1000 eV. Our MTCS values display the same disagreement with those of Ref.~\cite{McEachran2018Be} at very low energies as our ICS values did at the same energies.

\subsection{Magnesium}
\begin{figure*}
\includegraphics{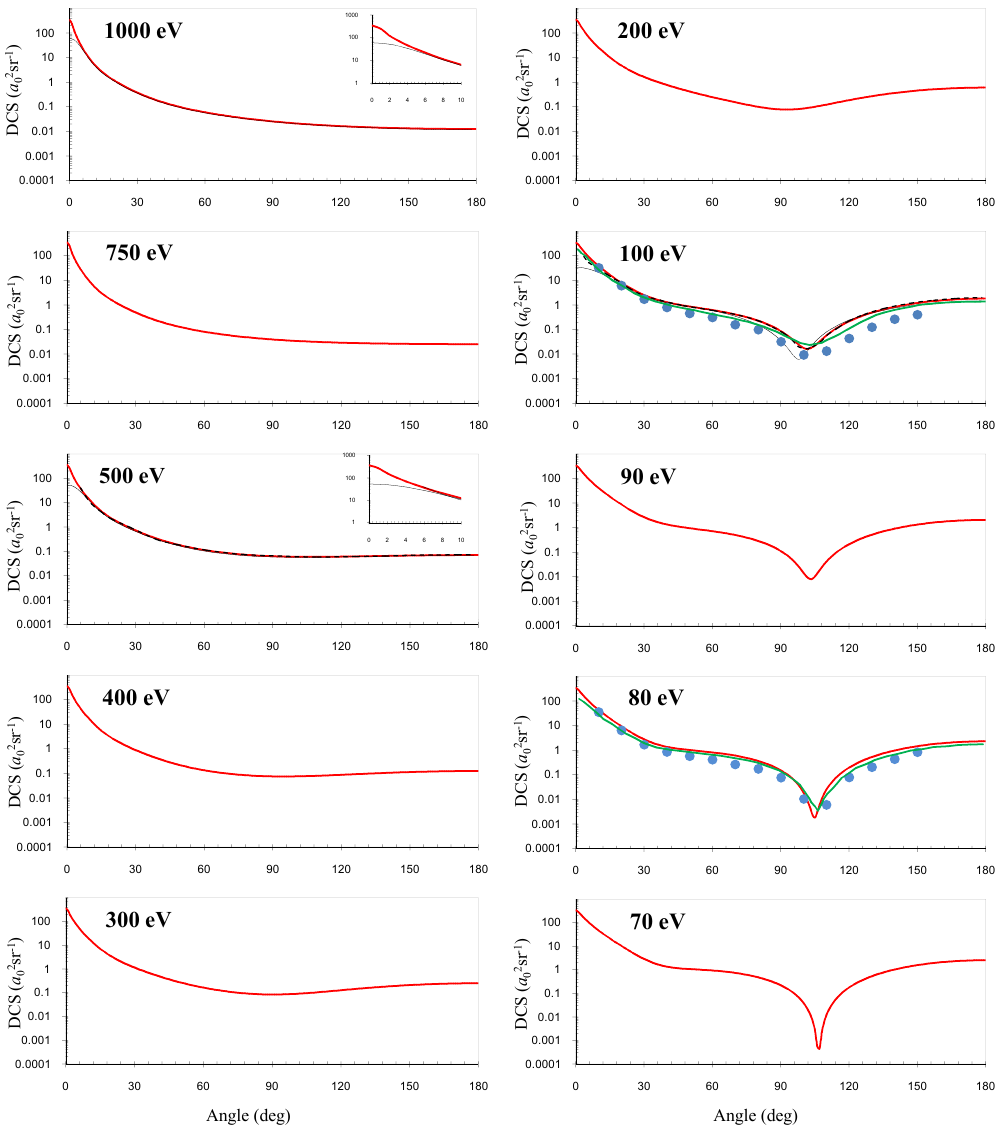}
\caption{\label{fig:mg1} Differential cross sections for elastic electron scattering from magnesium at energies between 70 and 1000 eV: thick solid red line, present work; Experimental data: {\color{cyan}$\mathlarger{\mathlarger{\mathlarger{\bullet}}}$}, Predojevi\'c \textit{et al.}~\cite{Predojevic2007}; Other theoretical: thin solid black line, Gregory and Fink~\cite{Gregory1974}; black dashed line, Khare \textit{et al.}~\cite{Khare1983}; thick green solid line, BSR-37 data of Zatsarinny \textit{et al.}~\cite{Zatsarinny2009}. Note that the size of the marker for the experimental data is indicative of its error.}
\end{figure*}

\begin{figure*}
\includegraphics{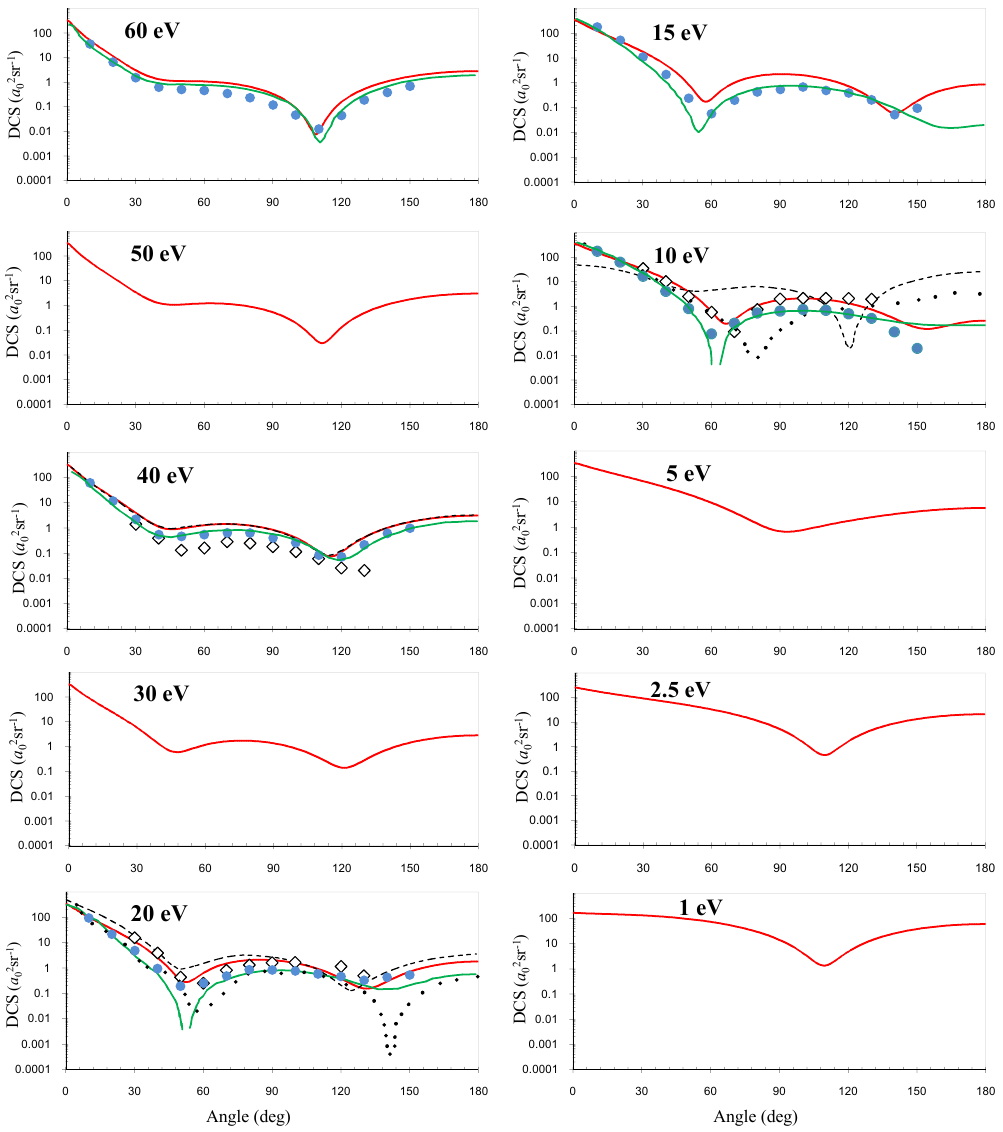}
\caption{\label{fig:mg2} Differential cross sections for elastic electron scattering from magnesium at energies between 1 and 60 eV: thick solid red line, present work; Experimental data: {\color{cyan}$\mathlarger{\mathlarger{\mathlarger{\bullet}}}$}, Predojevi\'c \textit{et al.}~\cite{Predojevic2007}; $\lozenge$, Williams and Trajmar~\cite{Williams1978}; Other theoretical: thin solid black line, Gregory and Fink~\cite{Gregory1974}; dotted black line, Fabrikant~\cite{Fabrikant1980}; black dashed line, Khare \textit{et al.}~\cite{Khare1983}; thick green solid line, BSR-37 data of Zatsarinny \textit{et al.}~\cite{Zatsarinny2009}. Note that the size of the marker for the experimental data is indicative of its error.}
\end{figure*}

We present our DCS results for magnesium in Figs.~\ref{fig:mg1} and \ref{fig:mg2}. Our DCS's for 100, 500, and 1000 eV are in good agreement with the results of Gregory and Fink~\cite{Gregory1974}, but the same disparity in the forward direction that we observed for beryllium in comparison of our results with those of Gregory and Fink~\cite{Fink1972}, exists here again since the  calculations of Gregory and Fink~\cite{Gregory1974} do not account for polarization and exchange; hence the same explanations for the disagreements in the forward direction given in the previous subsection apply here as well. At 100 eV impact energy, our DCS is in excellent agreement with experimental data of Predojevi\'c \textit{et al.}~\cite{Predojevic2007} in the forward direction but in the middle and backward scattering angles, though following congruently, our values generally stay above the experiment's. This gap is almost non-existent at 80 eV as our values and those of Zatsarinny \textit{et al.}~\cite{Zatsarinny2009} follow the experiment closely. Nonetheless, our DCS values indicate a slightly earlier minimum compared to those by Zatsarinny \textit{et al.}~\cite{Zatsarinny2009}, which is also visible at 100 eV. The disagreement on the positions of DCS's minima among most theoretical works is more evident in the comparisons in Fig.~\ref{fig:mg2}. As it can be seen in Fig.~\ref{fig:mg2}, while the agreement between our DCS's and the experimental data of Williams and Trajmar~\cite{Williams1978} at 10 and 20 eV are quite good, the comparison at 40 eV is not very promising. In fact, our DCS data visibly favors the experimental values of Predojevi\'c \textit{et al.}~\cite{Predojevic2007} at 40 eV.  In addition, the DCS's from the close-coupling calculations of Fabrikant~\cite{Fabrikant1980} at 10 and 20 eV do not agree very well, with those given by us and the experimental works. The DCS provided by the nonrelativistic partial-wave calculations of Khare \textit{et al.}~\cite{Khare1983} at 10 eV is, on the other hand, nowhere near any other available data for that energy. In all comparisons of our DCS data with the BSR-37 data of Zatsarinny \textit{et al.}~\cite{Zatsarinny2009} in Figs.~\ref{fig:mg1} and \ref{fig:mg2}, there exist disagreements on the positions of DCS minima. These discrepancies become more prominent at lower energies and at large scattering angles.

\begin{figure*}
\includegraphics{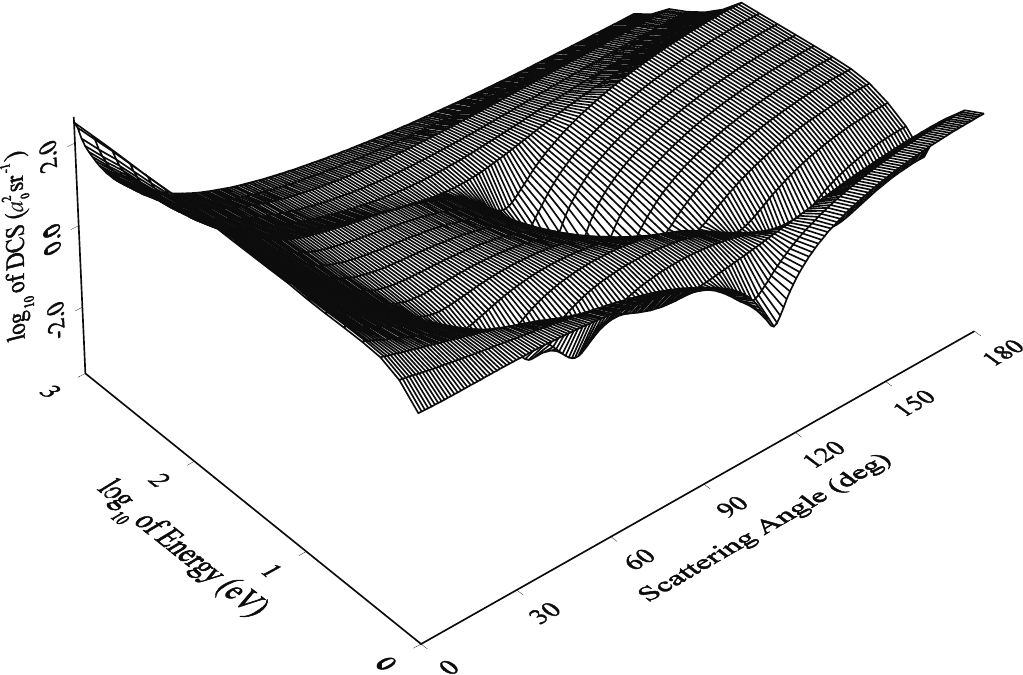}
\caption{\label{fig:mg3} a three-dimensional view of differential cross section for elastic electron scattering from magnesium.}
\end{figure*}

\begin{figure*}
\includegraphics{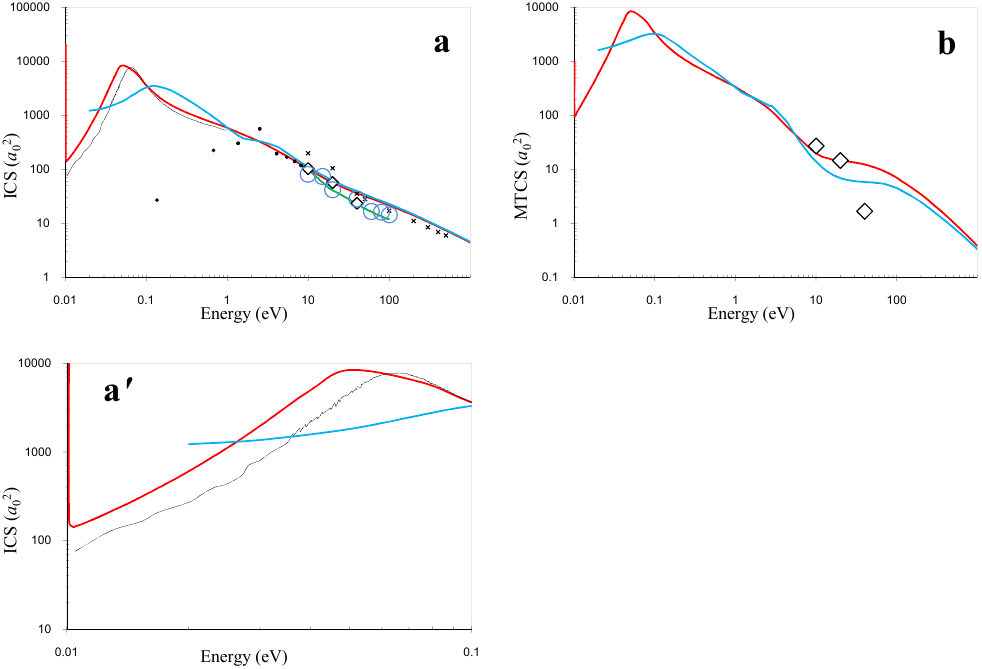}
\caption{\label{fig:mg4} Integrated (\textbf{a}, $\textbf{a}'$) and momentum transfer (\textbf{b}) cross sections for elastic electron scattering from magnesium: thick solid red line, present work; Experimental data: {\color{cyan}$\bigcirc$}, Predojevi\'c \textit{et al.}~\cite{Predojevic2007}; $\lozenge$, Williams and Trajmar~\cite{Williams1978}; Other theoretical: thin solid black line, Yuan and Zhang~\cite{Yuan1989}; $\times$, Khare \textit{et al.}~\cite{Khare1983}; $\mathsmaller{\bullet}$, Fabrikant~\cite{Fabrikant1975,Fabrikant1974}; thick blue solid line, McEachran \textit{et al.}~\cite{McEachran2018Mg}; thick green solid line, BSR-37 data of Zatsarinny \textit{et al.}~\cite{Zatsarinny2009}. Note that the size of the marker for the experimental data is indicative of its error.}
\end{figure*}

Like beryllium, a 3D graph of magnesium's DCS for energies between 1 and 1000 eV is given in Fig.~\ref{fig:mg3}. Magnesium's 3D graph demonstrates a more complex pattern compared to beryllium's, which reflects the difference in the atomic structure, e.g., the number of filled shells. In Fig.~\ref{fig:mg4}, we present our elastic ICS and MTCS results for magnesium. The agreement between our ICS and MTCS values and the experimental data of Williams and Trajmar~\cite{Williams1978} is satisfactory except for the data point of 40 eV, at which the DCS's also differed from each other in general but most notably in the backward direction. This difference can be explained knowing that the experimental ICS and MTCS of Ref.~\cite{Williams1978} have been obtained by the integration of the actual experimental DCS's and those obtained from the extrapolation to small and large scattering angles. The extrapolated DCS's for the large scattering angles, reported in Ref.~\cite{Williams1978}, do not bounce back and instead follow a downward path. Consequently, the discrepancy between our large-angle DCS values and the extrapolated DCS's of Williams and Trajmar~\cite{Williams1978} reach to an extreme, where ours is larger by a factor of 100. Obviously such discrepancy between the DCS's can cause a difference in ICS and MTCS values, since these quantities involve integrals of DCS over the angles. By the same reasoning we can see why there is a marginal agreement between the present ICS value at 40 eV and that of Predojevi\'c \textit{et al.}~\cite{Predojevic2007}.

Our ICS values are in very good agreement with those of McEachran \textit{et al.}~\cite{McEachran2018Mg} for energies between 5 and 1000 eV. That is also, more or less, the case for other theoretical values except those of Zatsarinny \textit{et al.}~\cite{Zatsarinny2009} which are consistently lower than our values. This is not surprising as Ref.~\cite{Zatsarinny2009}'s corresponding DCS values are also consistently smaller than ours over large intervals of scattering angles, if not all. Our low-energy ICS values are in agreement with the results of Yuan and Zhang~\cite{Yuan1989} down to 0.7 eV energy, below which our values are markedly higher. The ICS values of McEachran \textit{et al.}~\cite{McEachran2018Mg} deviate from our values below 4 eV and develop a maximum at almost 0.1 eV higher than ours. The present data observed a Ramsauer-Townsend (RT) minimum in the elastic ICS data for magnesium at just above 0.01 eV impact energy, below which the cross section rises very rapidly.

Our MTCS values are in excellent agreement with two experimental values of Williams and Trajmar~\cite{Williams1978} at 10 and 20 eV, however not at 40 eV as it was the case for the ICS data. The difference between our MTCS and the experimental MTCS at 40 eV is noticeably larger than the corresponding disparity in the ICS values. To understand this, one notices that the mathematical expression for MTCS has an extra factor of $\left( {1 - \cos \theta } \right)$ compared to that of ICS. This extra factor can contribute to this larger difference as it is small in the forward direction but becomes greater than 1 for $\theta > 90^ \circ$, where the present DCS values are maximally higher than those obtained by the extrapolation of the experimental DCS at 40 eV. The agreement with MTCS data of McEachran \textit{et al.}~\cite{McEachran2018Mg} is patchy. Angular dependence of DCS function at each projectile energy must be the contributing factor to the difference between our MTCS values and those of McEachran \textit{et al.}~\cite{McEachran2018Mg}. We observe a disagreement on the position of MTCS maximum (which existed in ICS data as well) between the only two theoretical values for magnesium.

\subsection{Calcium}
\begin{figure*}
\includegraphics{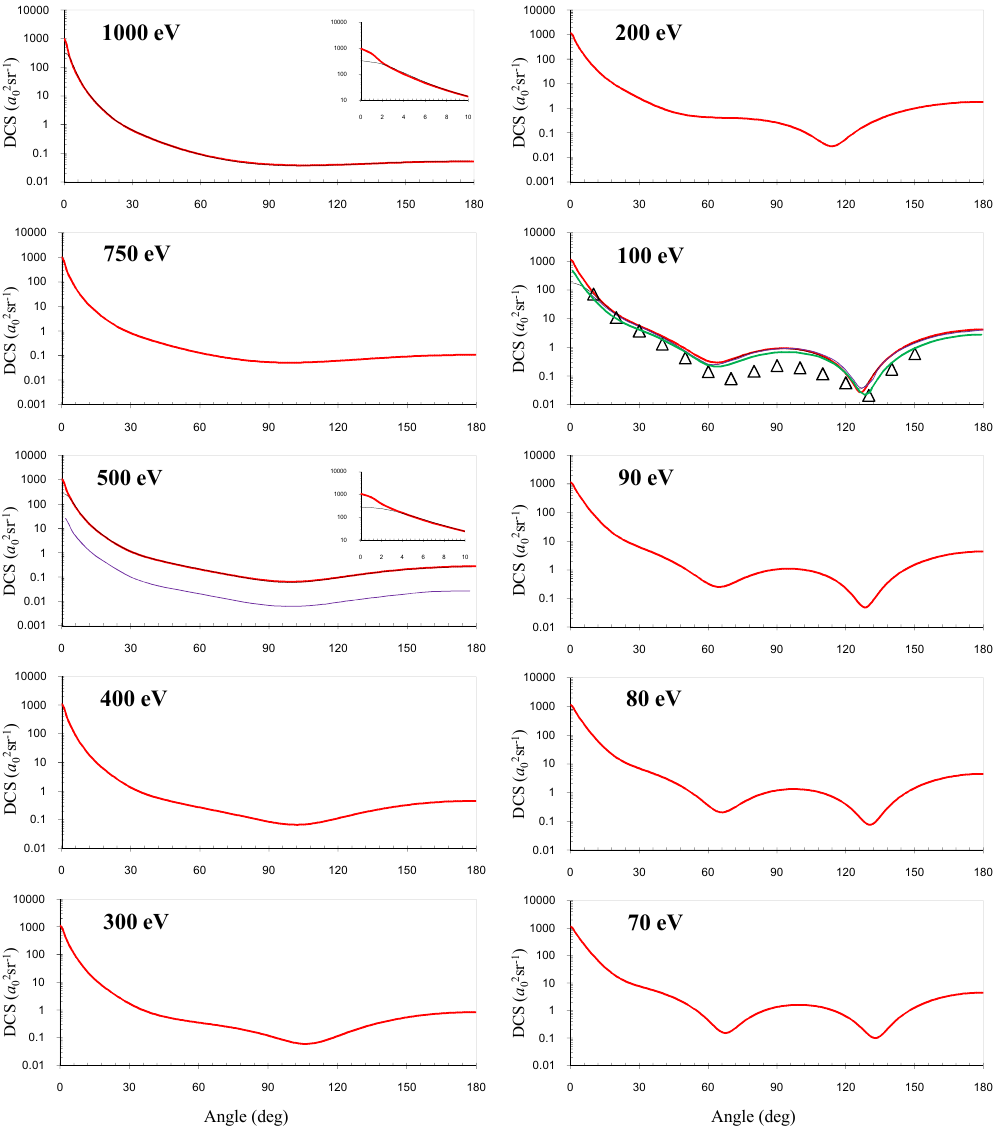}
\caption{\label{fig:ca1} Differential cross sections for elastic electron scattering from calcium at energies between 70 and 1000 eV: thick solid red line, present work; Experimental data: $\bigtriangleup$: Milisavljevi\'c \textit{et al.}~\cite{Milisavljevic2005}; Other theoretical: thin solid black line, Gregory and Fink~\cite{Gregory1974}; thin solid purple line, Khare \textit{et al.}~\cite{Khare1985}, thick solid green line, Zatsarinny \textit{et al.}~\cite{Zatsarinny2019}. Note that the size of the marker for the experimental data is indicative of its error.}
\end{figure*}

\begin{figure*}
\includegraphics{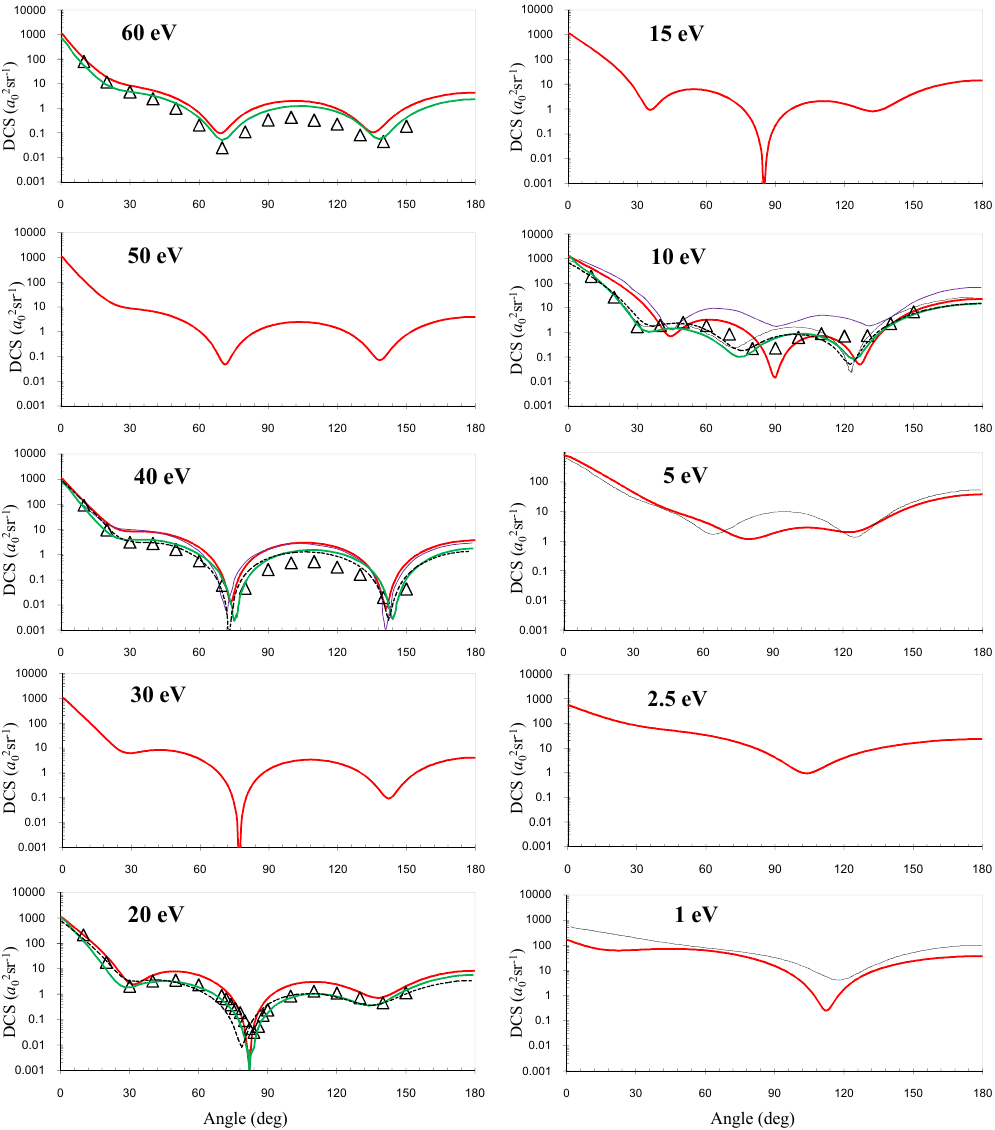}
\caption{\label{fig:ca2} Differential cross sections for elastic electron scattering from calcium at energies between 1 and 60 eV: thick solid red line, present work; Experimental data: $\bigtriangleup$, Milisavljevi\'c \textit{et al.}~\cite{Milisavljevic2005}; Other theoretical: thin solid black line, Yuan~\cite{Yuan1995}; thin solid purple line, Khare \textit{et al.}~\cite{Khare1985}, thick solid green line, BSR-438 data of Zatsarinny \textit{et al.}~\cite{Zatsarinny2019}, thick dashed black line, Wei \textit{et al.}~\cite{Wei2020}. Note that the size of the marker for the experimental data is indicative of its error.}
\end{figure*}

\begin{figure*}
\includegraphics{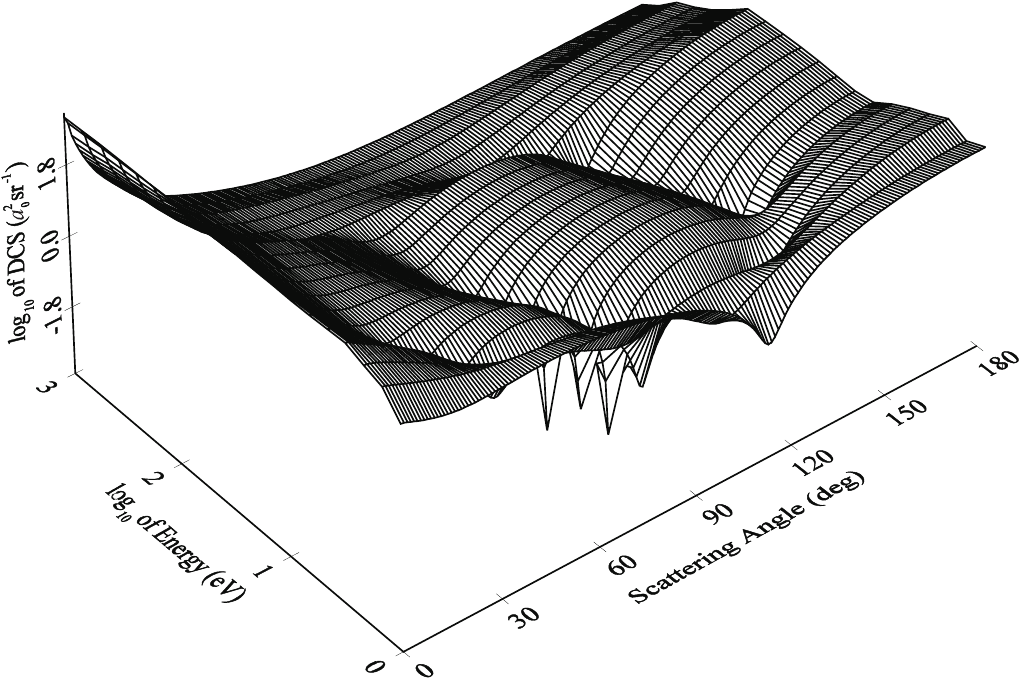}
\caption{\label{fig:ca3} a three-dimensional view of differential cross section for elastic electron scattering from calcium.}
\end{figure*}

Our DCS results, and comparison with other calculations, are shown in Figs.~\ref{fig:ca1} and \ref{fig:ca2}. They display noticeable diffraction oscillations, which are expected from a medium or heavy atom. These pronounced minima and their energy evolution can be seen readily in Fig.~\ref{fig:ca3}, where we present a 3D graph of calcium's DCS. As can be seen in Fig.~\ref{fig:ca1}, there is a good agreement between our DCS's and those of Gregory and Fink at 100, 500, and 1000 eV, while the aforementioned disparity in the forward direction still exists. The comparisons with the DCS's of Khare \textit{et al.}~\cite{Khare1985} show agreements at 40 and 100 eV, but strong disagreements at 10 and 500 eV impact energies. Our DCS values at 1, 5, and 10 eV also differ from the results of Yuan~\cite{Yuan1995}. Our DCS values are in very good agreement at 100 eV with those of Zatsarinny \textit{et al.}~\cite{Zatsarinny2019} but at 60, 40,  and 20 eV impact energies are generally larger and in mild disagreement on the position of the minima of DCS values with those of Ref.~\cite{Zatsarinny2019}. These disagreement with Ref.~\cite{Zatsarinny2019} are most apparent at 10 eV. The same disagreements exist with the DCS's of Wei \textit{et al.}~\cite{Wei2020} at 40, 20 and 10 eV. The comparisons of our DCS data with the only experimental work, by Milisavljevi\'c \textit{et al.}~\cite{Milisavljevic2005}, show general agreement in both magnitude and angular dependency and disagreement at 10 eV. Interestingly, similar discrepancies exist between the experimental values of Ref.~\cite{Milisavljevic2005} and other theoretical works. Further precision experimental investigations of elastic scattering of electrons off calcium atom would certainly be helpful.

\begin{figure*}
\includegraphics{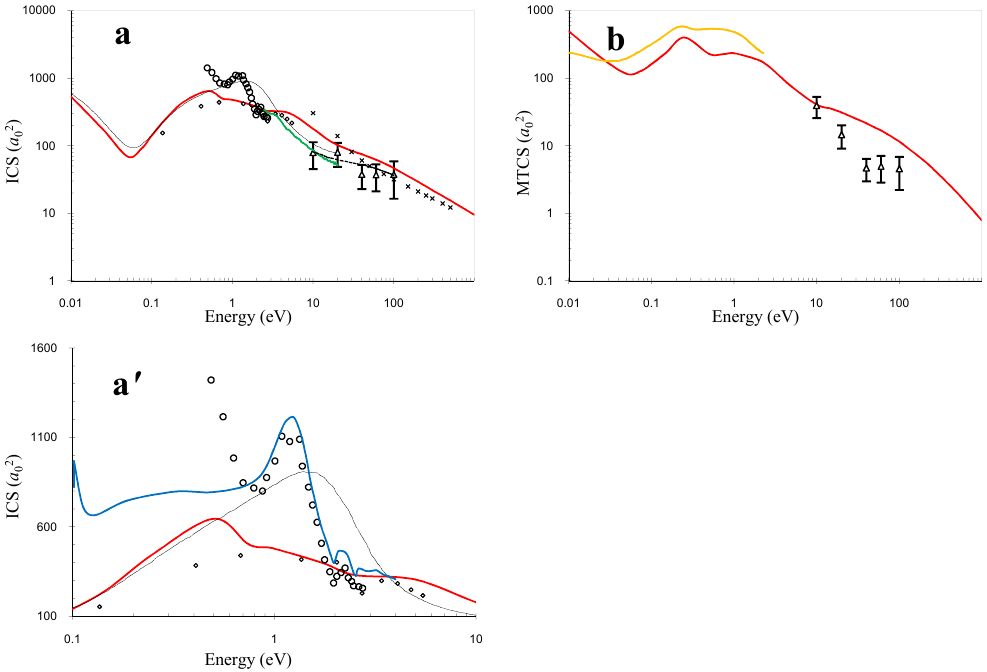}
\caption{\label{fig:ca4} Integrated (\textbf{a}, $\textbf{a}'$) and momentum transfer (\textbf{b}) cross sections for elastic electron scattering from calcium: thick solid red line, present work; Experimental data: $\circ$, Romanyuk \textit{et al.}~\cite{ Romanyuk1992}, $\mathsmaller{\bigtriangleup}$, Milisavljevi\'c \textit{et al.}~\cite{Milisavljevic2005}; Other theoretical: thin solid black line, Yuan~\cite{Yuan1995}; $\lozenge$, Fabrikant~\cite{Fabrikant1975}; $\times$, Khare \textit{et al.}~\cite{Khare1985}, thick solid green line, BSR-438 data of Zatsarinny \textit{et al.}~\cite{Zatsarinny2019}, thick solid blue line, BSR-39 data of Zatsarinny \textit{et al.}~\cite{Zatsarinny2006}, thick solid amber line, Cribakin \textit{et al.}~\cite{Cribakin1992}, thick dashed black line, Wei \textit{et al.}~\cite{Wei2020}.}
\end{figure*}

Our ICS and MTCS data for calcium are presented in Fig.~\ref{fig:ca4}. There is a partial agreement with the ICS values of Khare \textit{et al.}~\cite{Khare1985} for energies between 10 and 500 eV. Present ICS values demonstrate a good agreement with the ICS's of Yuan~\cite{Yuan1995} for energies below 0.5 eV, but differ from them strongly at larger energies. The ICS's of Fabrikant~\cite{Fabrikant1975} also disagree with those of Yuan~\cite{Yuan1995} at energies larger than 0.5 eV, while they somehow agree with ours. The experimental ICS values of Milisavljevi\'c \textit{et al.}~\cite{Milisavljevic2005} agree with our values at 100 and 20 eV while at other energies sit lower than ours. This is not surprising as our DCS values are also generally higher than those of Ref.~\cite{Milisavljevic2005}. Our TCS values are also higher than those of BSR-438 data of of Zatsarinny \textit{et al.}~\cite{Zatsarinny2019} between 3 and 20 eV and those of of Wei \textit{et al.}~\cite{Wei2020}.

The only experimental ICS data set at low energies~\cite{ Romanyuk1992} follows a complex path, which seems to prefer the low energy theoretical data of Zatsarinny \textit{et al.}~\cite{Zatsarinny2006} down to 0.7 eV before the points suddenly rise (which implies a relative minimum in ICS function). The estimated error in the experimental ICS values of Ref.~\cite{ Romanyuk1992} is not known to the authors and for that reason the size of the markers do not represent anything. Unfortunately, the ICS values of Zatsarinny \textit{et al.}~\cite{Zatsarinny2006} do not extend to lower energies to check the existence of an RT minimum, which our data and that of Yuan~\cite{Yuan1995} indicate at around 0.05 eV.

Our MTCS values follow a similar curve, for low impact energies, when compared to those of the theoretical work of Cribakin \textit{et al.}~\cite{Cribakin1992}, but they are generally lower than the values of Ref.~\cite{Cribakin1992} by, almost, a factor of two. Present MTCS values only agree with experimental work of Milisavljevi\'c \textit{et al.}~\cite{Milisavljevic2005} at 10 eV and not the rest. This agreement may be due to the excellent agreement of our DCS values in the backward direction with those of Milisavljevi\'c \textit{et al.}~\cite{Milisavljevic2005} at 10 eV, which is not replicated at other energies to that degree (see the last paragraph of the previous subsection for more on this).

\begin{figure*}
\includegraphics{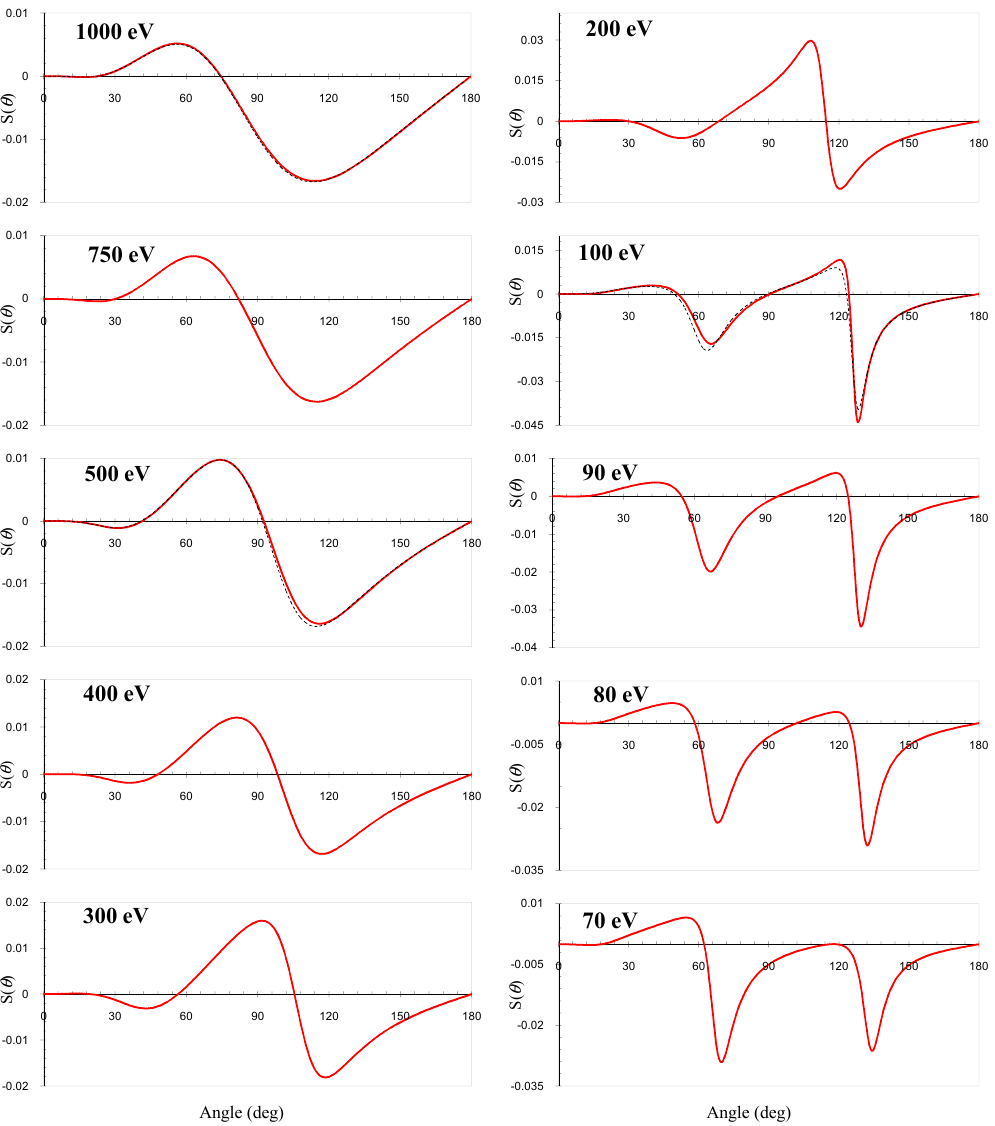}
\caption{\label{fig:ca5} Sherman functions for elastic electron scattering from calcium at energies between 70 and 1000 eV: thick solid line, present work; Other theoretical: dashed line, Gregory and Fink~\cite{Gregory1974}.}
\end{figure*}

\begin{figure*}
\includegraphics{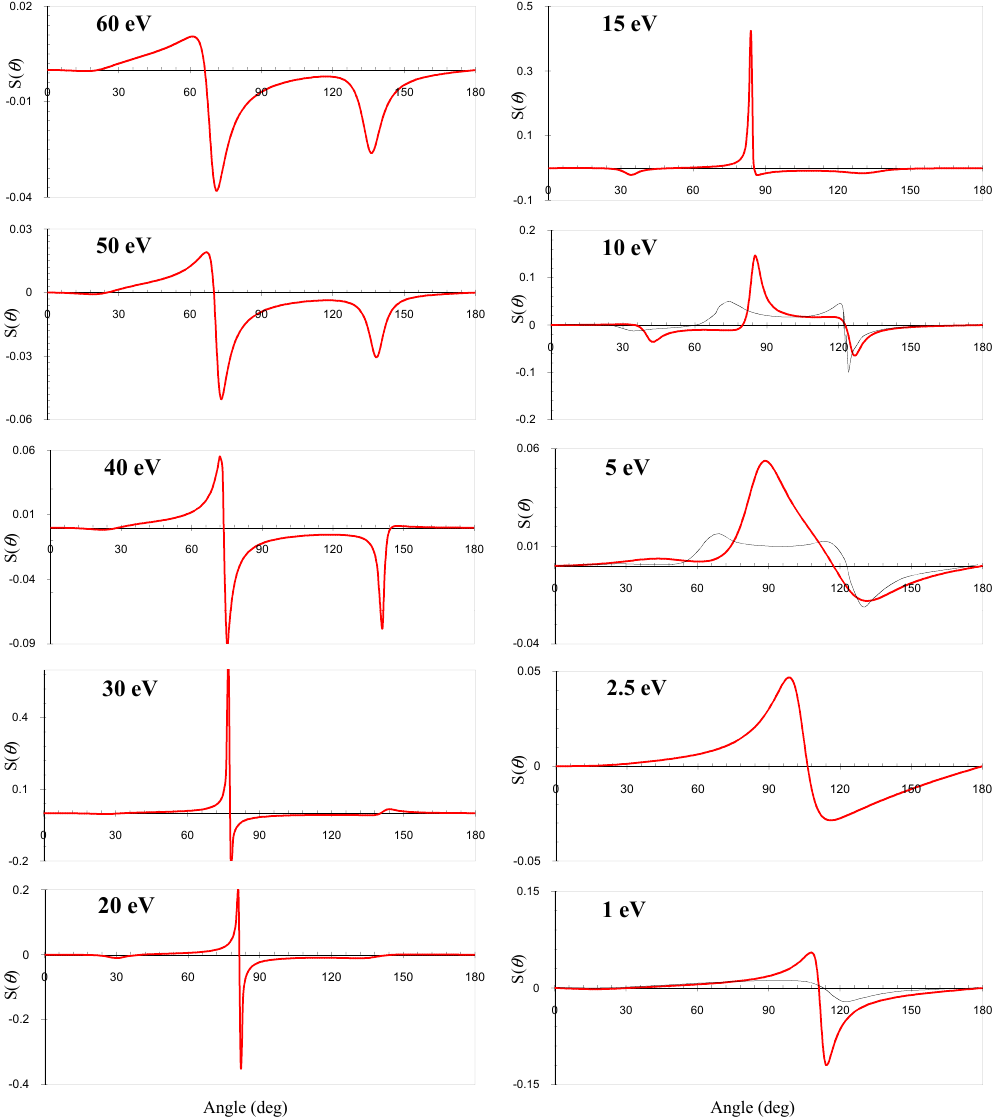}
\caption{\label{fig:ca6} Sherman functions for elastic electron scattering from calcium at energies between 1 and 60 eV: thick solid line, present work; Other theoretical: thin solid line, Yuan~\cite{Yuan1995}.}
\end{figure*}

\begin{figure*}
\includegraphics{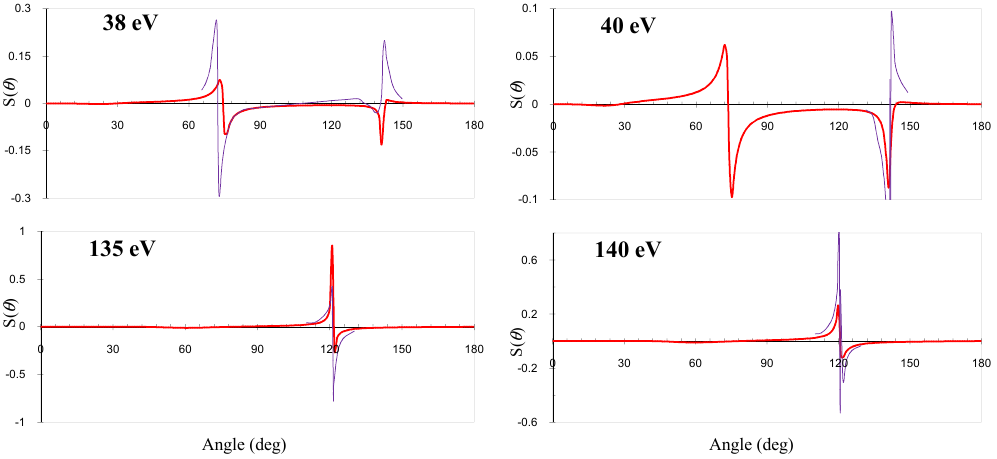}
\caption{\label{fig:ca7} Comparisons of Sherman functions for elastic electron scattering from calcium at various energies: thick solid line, present work; Other theoretical: dashed line, Khare \textit{et al.}~\cite{Khare1985}.}
\end{figure*}

In addition to cross sections, we also present Sherman functions for elastic electron scattering from calcium, in Figs.~\ref{fig:ca5} and \ref{fig:ca6}. The magnitude of Sherman function, $S(\theta)$, gives the degree of polarization of an unpolarized beam after scattering, given the direction of polarization being perpendicular to the scattering plane, the so-called up/down asymmetry. Furthermore, it can describe the left/right asymmetry in the scattering of a polarized beam. The sign of $S(\theta)$, on the other hand, indicates whether the spin-up states are preferentially populated, in which case it is positive, or the spin-down states are more occupied, in which case it is negative. The comparisons of our Sherman function results with those of Gregory and Fink~\cite{Gregory1974}, at 100, 500, and 1000 eV, indicate a very good agreement, while comparisons with the results of Yuan~\cite{Yuan1995} at 1, 5, and 10 eV show some disagreements. This is not surprising, however, if we recall the differences in the DCS's at those energies. In Fig.~\ref{fig:ca7}, we compare our Sherman functions with some of those given in the paper by Khare \textit{et al.}~\cite{Khare1985}. Generally, there is a mixed agreement between our data and those of Khare \textit{et al.}~\cite{Khare1985}. The Sherman functions provided by Ref.~\cite{Khare1985} do not span the whole angular range; therefore a thorough comparison is not possible. Even though Khare \textit{et al.}~\cite{Khare1985} do not give Sherman functions at 100 ad 500 eV, they mention a disagreement with the results of Gregory and Fink~\cite{Gregory1974} at those energies, with theirs being larger by a factor of two at all angles. Since our Sherman functions agree with those of Gregory and Fink~\cite{Gregory1974}, the Sherman functions of Ref.~\cite{Khare1985} at 100 and 500 eV, then, would have been larger than ours by a factor of two, as well.
 
\section{Conclusion}
We now have extended our method of calculating elastic electron scattering for inert gases to all alkaline-earth-metal atoms. Our DCS's compared well with available experimental DCS data, which were available only for magnesium and calcium. The present ICS values also produced agreements and disagreements with experimental and other theoretical results, such as B-spline, CCC and ROP calculations. We also studied the spin polarization of the scattered electrons from the calcium atom, and gave Sherman functions for an extensive energy range. Comparisons between our Sherman functions and prior theoretical results generated mixed outcome, which in the absence of experimental data or other precise calculations yield no final conclusion.


\end{document}